\documentclass[aps,pra,groupedaddress]{revtex4}%\usepackage{showkeys}
\usepackage{amsmath} 
\usepackage{amsfonts}
\usepackage{amstext}
\usepackage{latexsym, amssymb} 
\usepackage{bm} %per il grassetto
\usepackage[linktocpage=true,colorlinks=true,citecolor=blue]{hyperref} %funzia con sopra il dvips

\DeclareMathOperator{\Tr}{Tr} \makeatletter

\begin{document}
\title{\bf On the precise connection between the GRW master-equation
  and master-equations for the description of decoherence}
\author{Bassano \surname{Vacchini}}
\email{bassano.vacchini@mi.infn.it}
\affiliation{Dipartimento di Fisica
dell'Universit\`a di Milano, Via Celoria 16, 20133 Milan, Italy \\
Istituto Nazionale di Fisica Nucleare, sezione di Milano,
Via Celoria 16, 20133 Milan, Italy}
\begin{abstract}
   We point out that the celebrated GRW master-equation is invariant
   under translations, reflecting the homogeneity of space, thus
   providing a particular realization of a general class of
   translation-covariant Markovian master-equations. Such
   master-equations are typically used for the description of
   decoherence due to momentum transfers between system and
   environment. Building on this analogy we show the exact
   relationship between the GRW master-equation and decoherence
   master-equations, further providing a collisional decoherence model
   formally equivalent to the GRW master-equation.  This allows for a
   direct comparison of order of magnitudes of relevant parameters.
   This formal analogy should not lead to confusion on the utterly
   different spirit of the two research fields, in particular it has
   to be stressed that the decoherence approach does not lead to a
   solution of the measurement problem.  Building on this analogy
   however the feasibility of the extension of spontaneous
   localization models in order to avoid the infinite energy growth is
   discussed. Apart from a particular case considered in the paper,
   it appears that the amplification mechanism is generally spoiled by
   such modifications.
\end{abstract}
 \maketitle

\section{Introduction}
\label{sec:introduction}
The measurement problem in quantum mechanics has attracted and puzzled
physicists for decades, still remaining, together with the connected
issue of the relationship between quantum and classical world, one of
the main points of controversy, spurring further and deeper thinking
about the subject (see e.g.\cite{specialissue} for a most recent
collection of papers covering the subject from different
perspectives). An approach which has received considerable attention is
the one of dynamical reduction models\cite{GRW-review}, putting
forward stochastic and non--linear modifications of the Schr\"odinger
equation in order to reconcile microscopic and macroscopic
world. Techniques, ideas and equations used in dynamical reduction
models, and more generally for the study of the measurement problem,
are actually common to different other fields of physics, typically
open system theory\cite{Petruccione}, even if used in a utterly
different conceptual framework, so that these different lines of
research can benefit from each other. Not by chance the original
Ghirardi Rimini Weber (GRW)
model for spontaneous localization\cite{GRW} was actually inspired by
a seminal work on continuous quantum measurement\cite{continue1}
using similar tools with a quite different interpretation, as recently
stressed in a historical review of the spontaneous localization
approach to quantum mechanics\cite{RiminiAIP06}.

In the present paper we want to focus on the original GRW
master-equation, whose unravelling leads to a model of dynamical
reduction, showing that it can be rewritten in a simple way putting
into major evidence its basic features, and especially the fact that
it is a particular realization of the general class of
translation-covariant Markov master-equations described by
Holevo\cite{HolevoRMP32}. In such a way one can also easily introduce
a model of decoherence due to collisional interactions with a
background gas which would lead to a formally equivalent
master-equation, so that a direct comparison of the orders of
magnitude of the two different effects can be straightforwardly done,
obviously confirming the known estimates. A further advantage of this
approach is that one can now easily figure out a way to cope with the
problem of energy non conservation in the original GRW
model\cite{Ballentine-Gallis}. Extending the equivalent decoherence
model to describe also dissipation one indeed obtains a way to prevent
the energy from going to infinity. It appears however that in such a
way one of the basic features of the model, i.e. the increase of the
localization effect on the centre of mass of a composed system scaling
with the number of its constituents, sometimes called amplification
mechanism, is no more granted on general grounds. A notable exception
in this respect is the model considered in\cite{art11}, and we shall
clarify why it is so.

The paper is organized as follows: in
Sect.\ref{sec:transl-invar-struct} we show that the GRW
master-equation is a member of a general class of
translation-covariant master-equations, in
Sect.\ref{sec:equiv-decoh-model} building on the previous results we
introduce a simple decoherence model formally equivalent to the GRW
master-equation, finally in Sect.\ref{sec:energy-incr-spont} we point
out when the problem of energy non conservation can be solved in such
models, drawing conclusions in Sect.\ref{sec:conclusions-outlook}.

\section{Translational invariance and structure of the master-equation}
\label{sec:transl-invar-struct}

It is well--known that the GRW master-equation is closely related to
the master-equations used in decoherence models, so that
e.g. in\cite{KieferNEW} it is considered \textit{of the form} of the
Gallis and Fleming master-equation for the description of collisional
decoherence\cite{GallisPRA90}, and in\cite{TegmarkFPL93} it is argued
that \textit{scattering and the GRW effect have almost identical
  effects on the reduced density matrix}. We now want to fully clarify
and spell out in detail this relationship, showing that it is rooted
in a special property of the GRW master-equation, i.e. its
translation-covariance. Let us in fact call $\mathcal{L}_{\rm \scriptscriptstyle GRW}$ the
relevant part of the GRW master-equation\cite{GRW}, i.e. the
contributions apart from the Hamiltonian evolution
\begin{equation}
   \label{eq:1}
   \mathcal{L}_{\rm \scriptscriptstyle GRW}[\hat \rho]  
        =
        -\lambda
        \left \{\hat \rho -
         \left ({\frac{\alpha}{\pi}}\right )^\frac{3}{2}
        \int 
        d^3 \!
        {\bm{y}}
        \,  
         e^{-\frac{1}{2}\alpha ({\hat{\mathsf{x}}}-\bm{y})^2}
         \hat \rho
         e^{-\frac{1}{2}\alpha ({\hat{\mathsf{x}}}-\bm{y})^2}
        \right \},
\end{equation}
where ${\hat{\mathsf{x}}}$ and ${\hat{\mathsf{p}}}$ are
position and momentum operators of the particle subject to spontaneous
localization. One can immediately check that given the unitary representation ${\hat{\mathsf{U}}}
(\bm{a})=e^{{-\frac{i}{\hbar}\bm{a}\cdot{\hat{\mathsf{p}}}}}$,
$\bm{a}\in \mathbb{R}^3$, of the group of translations the following covariance equation\cite{HolevoNEW} is
satisfied thanks to the invariance under translations of the Lebesgue measure:
\begin{align}
   \label{eq:2}
      \mathcal{L}_{\rm \scriptscriptstyle GRW}[e^{{-\frac{i}{\hbar}\bm{a}\cdot{\hat{\mathsf{p}}}}}{\hat \rho}e^{{+\frac{i}{\hbar}\bm{a}\cdot{\hat{\mathsf{p}}}}}]
% &=        -\lambda
%         \left \{e^{{-\frac{i}{\hbar}\bm{a}\cdot{\hat{\mathsf{p}}}}}{\hat \rho} e^{{+\frac{i}{\hbar}\bm{a}\cdot{\hat{\mathsf{p}}}}}-
%          \left ({\frac{\alpha}{\pi}}\right )^\frac{3}{2}
%         \int 
%         d^3 \!
%         {\bm{y}}
%         \,  
%          e^{-\frac{1}{2}\alpha ({\hat{\mathsf{x}}}-\bm{y})^2}
%          e^{{-\frac{i}{\hbar}\bm{a}\cdot{\hat{\mathsf{p}}}}}{\hat \rho}e^{{+\frac{i}{\hbar}\bm{a}\cdot{\hat{\mathsf{p}}}}}
%          e^{-\frac{1}{2}\alpha ({\hat{\mathsf{x}}}-\bm{y})^2}
%         \right \}
% \nonumber
% \\
&=        -\lambda
e^{{-\frac{i}{\hbar}\bm{a}\cdot{\hat{\mathsf{p}}}}}
        \left \{\hat \rho -
         \left ({\frac{\alpha}{\pi}}\right )^\frac{3}{2}
        \int 
        d^3 \!
        {\bm{y}}
        \,  
         e^{-\frac{1}{2}\alpha ({\hat{\mathsf{x}}}+\bm{a}-\bm{y})^2}
         {\hat \rho}
         e^{-\frac{1}{2}\alpha ({\hat{\mathsf{x}}}+\bm{a}-\bm{y})^2}
        \right \}
e^{{+\frac{i}{\hbar}\bm{a}\cdot{\hat{\mathsf{p}}}}}
\nonumber
\\
&=e^{{-\frac{i}{\hbar}\bm{a}\cdot{\hat{\mathsf{p}}}}}\mathcal{L}_{\rm \scriptscriptstyle GRW}[{\hat \rho}]e^{{+\frac{i}{\hbar}\bm{a}\cdot{\hat{\mathsf{p}}}}}.
\end{align}
The action of the mapping $\mathcal{L}_{\rm \scriptscriptstyle GRW}$
giving the dynamics and the action
of the unitary representation of translations commute, reflecting the
invariance under translations of the underlying model. This is
actually a natural fact, since the modification of quantum mechanics
brought about by the GRW model would otherwise  break homogeneity of
space. In view of this property  it is quite natural to look
at~\eqref{eq:1} within the general characterization of
translation-covariant Markovian master-equations given by Holevo. For
the present discussion it is enough to consider the case of a bounded
mapping $\mathcal{L}$, so that its structure is given by\cite{HolevoRMP32}
\begin{equation}
   \label{eq:3}
      \mathcal{L}[{\hat \rho}]=
\int d\mu (\bm{q})\sum_{j=1}^{\infty}
\left[e^{{i\over\hbar}\bm{q}\cdot{\hat{\mathsf{x}}}}
L_j(\bm{q},{\hat{\mathsf{p}}})
{\hat \rho}
L^{\dagger}_j(\bm{q},{\hat{\mathsf{p}}})e^{-{i\over\hbar}\bm{q}\cdot{\hat{\mathsf{x}}}}
%\right.
%\\
%\left.
-        \frac 12
        \left \{
        L^{\dagger}_j(\bm{q},{\hat{\mathsf{p}}})L_j(\bm{q},{\hat{\mathsf{p}}}),{\hat \rho} 
        \right \}
\right],
\end{equation}
where $\bm{q}$ has the dimensions of momentum,
$L_j({\bm{q}},\cdot)$ are bounded functions, $\mu ({\bm{q}})$
is a positive $\sigma$-finite measure on $\mathbb{R}^3$ and $\int d\mu
({\bm{q}})
\sum_{j=1}^{\infty} |L_j({\bm{q}},\cdot)|^2<+\infty$. This is a general mathematical result, and
covariance under translations as in~\eqref{eq:2} can be easily
checked. 
To get a grasp on the physics that can be described by~\eqref{eq:3}
let us point out that the action of the unitary operators
$e^{{i\over\hbar}\bm{q}\cdot{\hat{\mathsf{x}}}}$ and
$e^{-{i\over\hbar}\bm{q}\cdot{\hat{\mathsf{x}}}}$, to the left and to the
right of the statistical operator respectively, corresponds to a momentum
transfer of amount $\bm{q}$, as can be easily seen from the fact that
\begin{equation}
   \label{eq:41}
    \langle \bm{p}|{\hat \varrho}|\bm{p}\rangle \longrightarrow \langle
 \bm{p}-\bm{q}|{\hat \varrho}|\bm{p}-\bm{q}\rangle 
\end{equation}
whenever
\begin{equation}
   \label{eq:45}
     {\hat \varrho} \longrightarrow e^{{i\over\hbar}\bm{q}\cdot {\hat
         {{x}}}}{\hat \varrho}e^{-{i\over\hbar}\bm{q}\cdot {\hat
         {{x}}}}.
\end{equation} 
The appearance of the $L_j(\bm{q},{\hat{\mathsf{p}}})$ operators
further implies that the momentum transferred to the massive particle
described by the statistical operator ${\hat \varrho}$ actually depends on the
momentum of the particle itself, so that effects like e.g. energy
relaxation can be described: depending on the value of its momentum
and therefore on its kinetic energy the particle gains or looses momentum and energy
in the single collision events. Of course in certain regimes this
dependence can be very weak, so that the momentum operator
${\hat{\mathsf{p}}}$ can be replaced by a reference value and the
mathematical structure of~\eqref{eq:3} simplifies a lot.

Let us now suppose in fact that the $L_j$ functions only depend on
$\bm{q}$, thus becoming $\mathbb{C}$-numbers instead of  operators. As
we shall see later on this missing of the ${\hat{\mathsf{p}}}$
dependence in the $L_j$ is strictly related to the infinite energy
growth in spontaneous localization dynamical reduction models. In view
of the previous requirements we can set
\begin{equation}
   \label{eq:4}
   \int d\mu
({\bm{q}})
\sum_{j=1}^{\infty} |L_j({\bm{q}})|^2 \equiv\lambda < +\infty,
\end{equation}
where $\lambda$ is a constant with dimensions of frequency, and
assuming $d\mu
({\bm{q}})$ absolutely continuous with respect to the Lebesgue
measure one can also set
\begin{equation}
   \label{eq:5}
d\mu
({\bm{q}})
\sum_{j=1}^{\infty} |L_j({\bm{q}})|^2 \equiv{\lambda}  d^3\!
        \bm{q}
        \,  \tilde{\mathcal{G}}^2 (\bm{q}),
\end{equation}
where without loss of generality we can take $\tilde{\mathcal{G}}(\bm{q})$
positive and such that its square $\tilde{\mathcal{G}}^2 (\bm{q})$ is
integrable over $L^2 (\mathbb{R}^3)$ and normalized to one, so that
$\tilde{\mathcal{G}}^2 (\bm{q})$ can be interpreted as a probability
density. In particular the positive function
$\tilde{\mathcal{G}}(\bm{q})$ can be seen as Fourier transform of a
function $\mathcal{G} (\bm{x})$ given by
\begin{equation}
   \label{eq:6}
   \mathcal{G} (\bm{x})=  \int \frac{d^3 \! \bm{q}}{(2\pi\hbar)^\frac{3}{2}} \,
e^{\frac{i}{\hbar}\bm{q}\cdot\bm{x}} \tilde{\mathcal{G}}(\bm{q}).
\end{equation}
% such that $\mathcal{G}^{\ast} (\bm{x})= \mathcal{G} (-\bm{x})$ and is
% positive definite in the sense that
% \begin{equation}
%    \label{eq:7}
%    \sum^{n}_{i,j=1} z_{i}^{\ast} \mathcal{G}
%    (\bm{x}_{i}-\bm{x}_{j})z_{j} \geq 0
% \end{equation}
% thanks to Bochner's theorem, in particular $\mathcal{G} (\bm{x})$ is a
% characteristic function\cite{FellerII}. 
In the considered case~\eqref{eq:3} highly simplifies and can be
written as
\begin{equation}
   \label{eq:8}
   \mathcal{L}[{\hat \rho}]=-\lambda
        \left \{\hat \rho -
\int
        d^3\!
        \bm{q}
        \,  \tilde{\mathcal{G}}^2 (\bm{q})
e^{{i\over\hbar}\bm{q}\cdot{\hat{\mathsf{x}}}}
{\hat \rho}
e^{-{i\over\hbar}\bm{q}\cdot{\hat{\mathsf{x}}}}
        \right \}, 
\end{equation}
so that by the positivity of $\mathcal{G} (\bm{x})$ one also has
\begin{align}
   \label{eq:9}
   \int
        d^3\!
        \bm{q}
        \,  \tilde{\mathcal{G}}^2 (\bm{q})
e^{{i\over\hbar}\bm{q}\cdot{\hat{\mathsf{x}}}}
{\hat \rho}
e^{-{i\over\hbar}\bm{q}\cdot{\hat{\mathsf{x}}}}
&=   \int
        d^3\!
        \bm{q}
        \,  
e^{{i\over\hbar}\bm{q}\cdot{\hat{\mathsf{x}}}}\tilde{\mathcal{G}} (\bm{q})
{\hat \rho}\tilde{\mathcal{G}} (\bm{q})
e^{-{i\over\hbar}\bm{q}\cdot{\hat{\mathsf{x}}}}
\nonumber\\
&=   \int
        d^3\!
        \bm{q} \int
        d^3\!
        \bm{k} \,   \delta^3 (\bm{k}-\bm{q})
e^{{i\over\hbar}\bm{q}\cdot{\hat{\mathsf{x}}}}\tilde{\mathcal{G}} (\bm{q})
{\hat \rho}\tilde{\mathcal{G}} (\bm{k})
e^{-{i\over\hbar}\bm{k}\cdot{\hat{\mathsf{x}}}}
\nonumber\\
&=   
\int
\frac{d^3 \! \bm{y}}{(2\pi\hbar)^{3}}
      \int  d^3\!\bm{q} \int
        d^3\!
        \bm{k} 
        \,  
e^{{i\over\hbar}\bm{q}\cdot ({\hat{\mathsf{x}}}-\bm{y})}\tilde{\mathcal{G}} (\bm{q})
{\hat \rho}\tilde{\mathcal{G}} (\bm{k})
e^{-{i\over\hbar}\bm{k}\cdot ({\hat{\mathsf{x}}}-\bm{y})}
\nonumber\\
&=   
\int
        d^3\!
        \bm{y} 
        \,  
\mathcal{G} ({\hat{\mathsf{x}}}-\bm{y})
{\hat \rho}
\mathcal{G}^{\dagger} ({\hat{\mathsf{x}}}-\bm{y})
\end{align}
and therefore~\eqref{eq:8} becomes
\begin{equation}
   \label{eq:10}
      \mathcal{L}[{\hat \rho}]=-\lambda
        \left \{\hat \rho -
\int
        d^3\!
        \bm{q}
        \,  \mathcal{G} ({\hat{\mathsf{x}}}-\bm{y})
{\hat \rho}
\mathcal{G}^{\dagger} ({\hat{\mathsf{x}}}-\bm{y})
        \right \}.
\end{equation}
It is now immediately apparent that the GRW master-equation is a
special case of~\eqref{eq:10} corresponding to the most natural choice
for the function $\tilde{\mathcal{G}} (\bm{q})$, i.e. a Gaussian
function, more precisely
\begin{equation}
   \label{eq:11}
   \tilde{\mathcal{G}}_{\rm \scriptscriptstyle GRW} (\bm{q})=
%\frac{1}{(\alpha\pi\hbar^2)^\frac{3}{2}}
\left       ({\frac{1}{\alpha\pi\hbar^2}}\right )^{3/2}
e^{-\frac{q^2}{2\alpha\hbar^2}}
\end{equation}
or equivalently
\begin{equation}
   \label{eq:12}
   \mathcal{G}_{\rm \scriptscriptstyle GRW} (\bm{x})=\left
      ({\frac{\alpha}{\pi}}\right )^{3/4}e^{-\frac{1}{2}\alpha x^2}
\end{equation}
where we have used the notation $q=|\bm{q}|$ and  $x=|\bm{x}|$,
so that~\eqref{eq:10} can be written in the two equivalent ways
\begin{align}
   \label{eq:13}
      \mathcal{L}_{\rm \scriptscriptstyle GRW}[\hat \rho]  
        &=
        -\lambda
        \left \{\hat \rho -
         \left ({\frac{\alpha}{\pi}}\right )^\frac{3}{2}
        \int 
        d^3 \!
        {\bm{y}}
        \,  
         e^{-\frac{1}{2}\alpha ({\hat{\mathsf{x}}}-\bm{y})^2}
         \hat \rho
         e^{-\frac{1}{2}\alpha ({\hat{\mathsf{x}}}-\bm{y})^2}
        \right \} 
\nonumber
\\
        &=
        -\lambda
\left \{\hat \rho -
\left({\frac{1}{\alpha\pi\hbar^2}}\right )^{3/2}
%        \left \{\hat \rho -\frac{1}{(\alpha\pi\hbar^2)^\frac{3}{2}}
        \int 
        d^3 \!
        {\bm{q}}
        \,  
e^{-\frac{q^2}{2\alpha\hbar^2}}
 e^{{i\over\hbar}\bm{q}\cdot{\hat{\mathsf{x}}}}
{\hat \rho}
e^{-{i\over\hbar}\bm{q}\cdot{\hat{\mathsf{x}}}}
        \right \} .
\end{align}
While the first line corresponds to the usual way of writing the
equation, the second line puts immediately into evidence the
connection with models of decoherence due to momentum transfer
events\cite{AlickiPRA02,KieferNEW,vienna,art12}. 
Note that the matrix elements in
the position representation of~\eqref{eq:8} have the general form
\begin{equation}
   \label{eq:14}
   \langle\bm{x}|\mathcal{L}[\hat \rho]|\bm{y}\rangle
=-\lambda
        \left \{1 -
\int
        d^3\!
        \bm{q}
        \,  \tilde{\mathcal{G}}^2 (\bm{q})
e^{{i\over\hbar}\bm{q}\cdot (\bm{x}-\bm{y})}
\right\}
\langle\bm{x}|{\hat \rho}|\bm{y}\rangle,
\end{equation}
where due to the fact that $\tilde{\mathcal{G}}^2 (\bm{q})$ is a
probability density its Fourier transform actually is a characteristic
function, with all its important properties\cite{FellerII,art12}. In
particular due to the fact that the Fourier transform of a product is
mapped into a convolution one has
\begin{equation}
   \label{eq:15}
   \int
        d^3\!
        \bm{q}
        \,  \tilde{\mathcal{G}}^2 (\bm{q})
e^{{i\over\hbar}\bm{q}\cdot \bm{x}}= (\mathcal{G}\ast \mathcal{G}) (\bm{x})
\end{equation}
and therefore
\begin{equation}
   \label{eq:16}
      \langle\bm{x}|\mathcal{L}[\hat \rho]|\bm{y}\rangle
=-\lambda
        \left \{1 -(\mathcal{G}\ast \mathcal{G}) (\bm{x}-\bm{y})
\right\}
\langle\bm{x}|{\hat \rho}|\bm{y}\rangle,
\end{equation}
as can be obtained directly from~\eqref{eq:10}. In the particular case
of a Gaussian function, the convolution again leads to a Gaussian
function, $(\mathcal{G}_{\rm \scriptscriptstyle GRW}\ast
\mathcal{G}_{\rm \scriptscriptstyle GRW})
(\bm{x})=e^{-\frac{\alpha}{4}x^2}$, 
% \begin{equation}
%    \label{eq:17}
%    (\mathcal{G}_{\rm \scriptscriptstyle GRW}\ast \mathcal{G}_{\rm \scriptscriptstyle GRW}) (\bm{x})=e^{-\frac{\alpha}{4}x^2}
% \end{equation}
and the matrix elements of the second line of~\eqref{eq:13}
immediately give the correct result
\begin{align}
   \label{eq:18}
    \langle\bm{x}|\mathcal{L}_{\rm \scriptscriptstyle GRW}[\hat \rho]|\bm{y}\rangle
&=-\lambda
        \left \{1 -(\mathcal{G}_{\rm \scriptscriptstyle GRW}\ast \mathcal{G}_{\rm \scriptscriptstyle GRW}) (\bm{x}-\bm{y})
\right\}
\langle\bm{x}|{\hat \rho}|\bm{y}\rangle
\nonumber
\\
&=-\lambda
        \left \{1 -e^{-\frac{\alpha}{4} (\bm{x}-\bm{y})^2}
\right\}
\langle\bm{x}|{\hat \rho}|\bm{y}\rangle.
\end{align}
Note that in this particular case the characteristic function is real
corresponding to the fact that the Gaussian has zero mean. The GRW
master-equation is thus a possible realization of the class of
translation-covariant master-equations given by~\eqref{eq:10}, each
uniquely characterized by a rate $\lambda$ and the choice of a
probability density $\tilde{\mathcal{G}}^2 (\bm{q})$. A general analysis of master-equations
like~\eqref{eq:10} and generalizations thereof in view of a
theoretical description of decoherence, also in connection with
experiments, has been done in\cite{art12}, to which we refer the
reader for further details. 

A key feature of the GRW model, i.e. the amplification mechanism, is
common to all choices of probability density, as follows immediately
from the fact that the position operator only appears in the unitary
operators $e^{{i\over\hbar}\bm{q}\cdot{\hat{\mathsf{x}}}}$ and their
adjoints. Suppose in fact to consider a system of $N$ particles,
of one type for the sake of simplicity, so that the dynamics would be
given by
\begin{equation}
   \label{eq:19}
   \mathcal{L}_{\rm \scriptscriptstyle GRW}[\hat \rho_{\rm \scriptscriptstyle tot}]=\sum_{i=1}^{N}  \mathcal{L}_{\rm \scriptscriptstyle GRW}^{i}[\hat \rho_{\rm \scriptscriptstyle tot}]
\end{equation}
where $\hat \rho_{\rm \scriptscriptstyle tot}$ is the $N$ particle statistical operator and
\begin{equation}
   \label{eq:20}
   \mathcal{L}_{\rm \scriptscriptstyle GRW}^{i}[\hat \rho_{\rm \scriptscriptstyle tot}]=-\lambda
        \left \{\hat \rho_{\rm \scriptscriptstyle tot} -
\int
        d^3\!
        \bm{q}
        \,  \tilde{\mathcal{G}}_{\rm \scriptscriptstyle GRW}^2 (\bm{q})
e^{{i\over\hbar}\bm{q}\cdot{\hat{\mathsf{x}}}_{i}}
{\hat \rho}_{\rm \scriptscriptstyle tot}
e^{-{i\over\hbar}\bm{q}\cdot{\hat{\mathsf{x}}}_{i}}
        \right \}, 
\end{equation}
where ${\hat{\mathsf{x}}}_{i}$ is the position operator of the $i$-th
particle. Switching to centre of mass coordinates with the linear transformation
\begin{equation}
   \label{eq:21}
   \bm{r}_i=\sum_{k=1}^{N}\Lambda_{ik}\bm{x}_k
\end{equation}
with $\Lambda_{1i}=\frac{m_i}{M}$ ($M=\sum_{i=1}^{N} m_i$), so that
\begin{equation}
   \label{eq:22}
   \bm{r}_1=\sum_{i=1}^{N}\frac{m_i}{M}\bm{x}_i\equiv \bm{X},
\end{equation}
where $\bm{X}$ denotes the coordinate of the centre of mass, one
immediately has
\begin{equation}
   \label{eq:23}
   \bm{x}_i=\bm{X}+\sum_{k=2}^{N}\Lambda_{ik}^{-1}\bm{r}_k
\end{equation}
and considering the partial trace with respect to the relative
coordinates $\bm{r}_2, \ldots,\bm{r}_N$ one has
\begin{equation}
   \label{eq:24}
   {\hat \rho}_{\rm \scriptscriptstyle CM}=\Tr_{\rm \scriptscriptstyle rel} {\hat \rho}_{\rm \scriptscriptstyle tot}
\end{equation}
and exploiting the properties of the trace operation
\begin{align}
   \label{eq:25}
   \Tr_{\rm \scriptscriptstyle rel}\mathcal{L}_{\rm \scriptscriptstyle GRW}^{i}[\hat \rho_{\rm \scriptscriptstyle tot}]
&=
-\lambda
        \left \{\hat \rho_{\rm \scriptscriptstyle CM} -
\int
        d^3\!
        \bm{q}
        \,  \tilde{\mathcal{G}}_{\rm \scriptscriptstyle GRW}^2 (\bm{q})
\Tr_{\rm \scriptscriptstyle rel}
\left(
e^{{i\over\hbar}\bm{q}\cdot (\hat{\mathsf{X}}+\sum_{k=2}^{N}\Lambda_{ik}^{-1}\hat{\mathsf{r}}_k) }
{\hat \rho}_{\rm \scriptscriptstyle tot}
e^{-{i\over\hbar}\bm{q}\cdot (\hat{\mathsf{X}}+\sum_{k=2}^{N}\Lambda_{ik}^{-1}\hat{\mathsf{r}}_k) }
\right)      
  \right \}
\nonumber \\
&=
-\lambda
        \left \{\hat \rho_{\rm \scriptscriptstyle CM} -
\int
        d^3\!
        \bm{q}
        \,  \tilde{\mathcal{G}}_{\rm \scriptscriptstyle GRW}^2 (\bm{q})
\Tr_{\rm \scriptscriptstyle rel}
\left(
e^{{i\over\hbar}\bm{q}\cdot \sum_{k=2}^{N}\Lambda_{ik}^{-1}\hat{\mathsf{r}}_k }
e^{{i\over\hbar}\bm{q}\cdot \hat{\mathsf{X}} }
{\hat \rho}_{\rm \scriptscriptstyle tot}
e^{-{i\over\hbar}\bm{q}\cdot \hat{\mathsf{X}} }
e^{-{i\over\hbar}\bm{q}\cdot\sum_{k=2}^{N}\Lambda_{ik}^{-1}\hat{\mathsf{r}}_k }
\right)      
  \right \}
\nonumber \\
&=
-\lambda
        \left \{\hat \rho_{\rm \scriptscriptstyle CM} -
\int
        d^3\!
        \bm{q}
        \,  \tilde{\mathcal{G}}_{\rm \scriptscriptstyle GRW}^2 (\bm{q})
e^{{i\over\hbar}\bm{q}\cdot \hat{\mathsf{X}} }
\hat \rho_{\rm \scriptscriptstyle CM}
e^{-{i\over\hbar}\bm{q}\cdot \hat{\mathsf{X}} }
  \right \},
\end{align}
so that one has a closed equation for the centre of mass degrees of
freedom with an equation of the same form apart from a rescaled
frequency $\lambda_N=N\lambda$
\begin{equation}
   \label{eq:26}
   \mathcal{L}_{\rm \scriptscriptstyle GRW}[\hat \rho_{\rm \scriptscriptstyle CM}]=-N\lambda\left \{\hat \rho_{\rm \scriptscriptstyle CM} -
\int
        d^3\!
        \bm{q}
        \,  \tilde{\mathcal{G}}_{\rm \scriptscriptstyle GRW}^2 (\bm{q})
e^{{i\over\hbar}\bm{q}\cdot \hat{\mathsf{X}} }
\hat \rho_{\rm \scriptscriptstyle CM}
e^{-{i\over\hbar}\bm{q}\cdot \hat{\mathsf{X}} }
  \right \}.
\end{equation}
This result ensures the amplification mechanism.

\section{Equivalent decoherence model}
\label{sec:equiv-decoh-model}

As explained in the previous section the GRW master-equation due to
the property of covariance under translations shares the same
structure as many models used for the description of decoherence. We
now briefly point out a model of collisional decoherence which would
lead to the very same master-equation, thus allowing for a direct
comparison of the order of magnitudes  of the GRW effect and of
collisional decoherence, as well as preparing for the discussion of
how to face the infinite energy growth. For the case of a massive
particle interacting through collisions with a free non degenerate gas
the master-equation can be written as
\begin{equation}
   \label{eq:27}
   \mathcal{L}_{\rm \scriptscriptstyle coll}[{\hat \rho}]= {2\pi \over\hbar}
        (2\pi\hbar)^3
        n
        \int d^3\!
        \bm{q}
        \,  
        {
        | \tilde{t} (q) |^2
        }
        \Biggl[
        e^{{i\over\hbar}\bm{q}\cdot{\hat{\mathsf{x}}}}
        \sqrt{
        S(\bm{q},E (\bm{q},{\hat{\mathsf{p}}}))
        }
        {\hat \rho}
        \sqrt{
        S(\bm{q},E (\bm{q},{\hat{\mathsf{p}}}))
        }
        e^{-{i\over\hbar}\bm{q}\cdot{\hat{\mathsf{x}}}}
        -
        \frac 12
        \left \{
        S(\bm{q},E (\bm{q},{\hat{\mathsf{p}}})),
        {\hat \rho}
        \right \}
        \Biggr],
\end{equation}
where $\tilde{t} (q) $ is the Fourier transform of the two-body
interaction potential between test particle and gas particles, $n$ the
density of the homogeneous background gas and $ S(\bm{q},E
(\bm{q},{\hat{\mathsf{p}}}))$ a positive two-point correlation
function, depending on momentum transfer $\bm{q}$ and energy transfer $E
(\bm{q},{\hat{\mathsf{p}}})=\frac{q^2}{2M}+\frac{{\hat{\mathsf{p}}}\cdot\bm{q}}{2M}$
($M$ being the mass of the test particle), known
as dynamic structure factor, accounting for the properties of the gas
(see\cite{art3,art5,art4,art10} for a reference
and\cite{HornbergerPRL06} for further extensions). The dynamic
structure factor accounts for energy and momentum transfer between
test particle and gas and for a free gas of Maxwell-Boltzmann
particles can be written
\begin{equation}
   \label{eq:28}
            S_{\rm \scriptscriptstyle MB}(\bm{q},E)
        =
        \sqrt{\frac{\beta m}{2\pi}}        
        {
        1
        \over
        q
        }
       e^{-{
        \beta
        \over
             8m
        }
        {
        (2mE + q^2)^2
        \over
                  q^2
        }}
\end{equation}
with $\beta$ the inverse temperature and $m$ the mass of the gas
particles. 
Neglecting in the first instance the energy dependence, related to
dissipation and evaluating the dynamic structure factor for zero
energy transfer one has
\begin{equation}
   \label{eq:29}
     \mathcal{L}_{\rm \scriptscriptstyle coll}[{\hat \rho}] = {2\pi \over\hbar}
        (2\pi\hbar)^3
        n
 \sqrt{\frac{\beta m}{2\pi}}  
        \int d^3\!
        \bm{q}
        \,  
        \frac{{
        | \tilde{t} (q) |^2
        }}{q}
      e^{-\frac{\beta}{8m}q^2}
              \Biggl[
        e^{{i\over\hbar}\bm{q}\cdot{\hat{\mathsf{x}}}}
        {\hat \rho}
        e^{-{i\over\hbar}\bm{q}\cdot{\hat{\mathsf{x}}}}
-
        {\hat \rho}
        \Biggr] .
\end{equation}
This expression is exactly of the same form as the GRW master-equation
if the interaction potential $t (x)$ is proportional to
$\frac{1}{x^{7/2}}$, so that according to the formula\cite{GelfandShilov}
\begin{equation}
   \label{eq:30}
   \int
        d^3\!
        \bm{x}
        \,  
e^{-{i\over\hbar}\bm{q}\cdot\bm{x}} x^{\mu}=2^{\mu+\frac{3}{2}}
(2\pi)^\frac{3}{2}\frac{\Gamma\left(
     \frac{\mu}{2}+\frac{3}{2}\right)}{\Gamma\left(
        -\frac{\mu}{2} \right) }\left( \frac{q}{\hbar}\right)^{-\mu-3}
\end{equation}
one has for  $t (x)=\frac{K}{x^{7/2}}$, with $K$ a coupling constant
\begin{equation}
   \label{eq:31}
   \tilde{t} (q)=\int \frac{d^3 \! \bm{x}}{(2\pi\hbar)^\frac{3}{2}} \,
e^{-\frac{i}{\hbar}\bm{q}\cdot\bm{x}} t (x)=-\frac{4}{3}\frac{K (2\pi)^\frac{3}{2}}{(2\pi\hbar)^{3}}\left( \frac{q}{\hbar}\right)^{1/2}
\end{equation}
and therefore
\begin{equation}
   \label{eq:32}
   \mathcal{L}_{\rm \scriptscriptstyle coll}[{\hat \rho}] =       -\lambda_{\rm \scriptscriptstyle coll}
        \left \{\hat \rho -
%\frac{1}{(\alpha_{\rm \scriptscriptstyle coll}\pi\hbar^2)^\frac{3}{2}}
\left       ({\frac{1}{\alpha_{\rm \scriptscriptstyle coll}\pi\hbar^2}}\right )^{3/2}
        \int 
        d^3 \!
        {\bm{q}}
        \,  
e^{-\frac{q^2}{2\alpha_{\rm \scriptscriptstyle coll}\hbar^2}}
 e^{{i\over\hbar}\bm{q}\cdot{\hat{\mathsf{x}}}}
{\hat \rho}
e^{-{i\over\hbar}\bm{q}\cdot{\hat{\mathsf{x}}}}
        \right \},
\end{equation}
with constants given by
\begin{equation}
   \label{eq:33}
   \alpha_{\rm \scriptscriptstyle coll}=\frac{16\pi}{(2\pi\beta\hbar^2)/m}=\frac{16\pi}{\lambda^2_{\rm \scriptscriptstyle th}}
\end{equation}
with ${\lambda_{\rm \scriptscriptstyle th}}$ the thermal wavelength of the gas particles, and
\begin{equation}
   \label{eq:34}
   \lambda_{\rm \scriptscriptstyle coll}=nm\frac{16\pi}{\lambda^2_{\rm
       \scriptscriptstyle th}} \frac{8}{9}\left(
      \frac{2\pi}{\hbar}\right)^{3}\frac{|K|^2}{\pi}=nm\, \alpha_{\rm \scriptscriptstyle coll} \frac{8}{9}\left( \frac{2\pi}{\hbar}\right)^{3}\frac{|K|^2}{\pi},
\end{equation}
so that as expected the key ingredients are the thermal wavelength,
the particle density and the strength of the interaction potential,
determining the scattering cross-section: the coherence length
${\lambda_{\rm \scriptscriptstyle th}}$ of the gas sets the scale of
the localization and the scattering cross-section fixes the frequency
of the localization events. Since the choice of the
interaction potential was essentially aimed at providing a possible
decoherence model which exactly reproduces the master-equation in the
GRW model, only order of magnitudes are here of relevance and typical
possible values of $K$\cite{Maitland} lead to a product
$\alpha_{\rm \scriptscriptstyle coll}\lambda_{\rm \scriptscriptstyle
  coll}$ 
%of the order of $10^{38}\, \mathrm{m}^{-2}\mathrm{sec}^{-1}$, much 
stronger
by orders of magnitude
than the GRW effect (see the interesting work of
Tegmark\cite{TegmarkFPL93}, also summarized in\cite{KieferNEW}
and\cite{GRW-review} for a more detailed analysis of orders of
magnitudes).

\section{Energy increase in spontaneous localization dynamical reduction models}
\label{sec:energy-incr-spont}

In the previous paragraph we have briefly introduced a collisional
decoherence model which exactly reproduces the GRW master-equation
with new parameters $\alpha_{\rm \scriptscriptstyle coll}$ and
$\lambda_{\rm \scriptscriptstyle coll}$ fixed by the bath properties.
Exploiting this formal correspondence one can easily figure out how to
cure the infinite energy growth common to both dynamical reduction
models and decoherence models. From the standpoint of particle gas
interaction this drawback is due to the fact that in decoherence
models energy transfers between particle and bath, leading to
dissipative effects, are simply not described. One therefore simply
has to look for an extension of \eqref{eq:29} including energy
relaxation. Such an equation is the natural counterpart of the
classical linear Boltzmann equation and is given by \eqref{eq:27} when
the dependence on the energy transfer in the dynamic structure factor
is not neglected
(see\cite{DiosiEL95,art3,art5,art4,art10,ZeilingerQBM-th,HornbergerPRL06}
for a more detailed treatment). It still complies with
translation-covariance and is in fact a possible realization
of~\eqref{eq:3} when the functions $L_j$ also depend on the momentum
operator of the particle. It would therefore be quite natural to build
on such a model to propose, with a suitable interpretation of the
parameters, a master-equation for a dynamical reduction model leading
to a finite energy value, further using a suitable unravelling
preserving the localization features. A non trivial difficulty however
appears, which was already encountered in early attempts to find
alternative model to the GRW master-equation\cite{BenattiNC}, and is
here strengthened by the available characterization of
translation-covariant master-equations given by
Holevo\cite{HolevoRMP32,HolevoRMP33}. When the functions $L_j$
actually depend on the momentum ${\hat{\mathsf{p}}}$ the amplification
mechanism is generally no more available, apart from very particular
cases which we are now in the position to spell out. An interesting
possibility appears to be the one considered in\cite{art11}, in which
the master-equation associated to dynamical reduction is the quantum
counterpart of the classical Fokker-Planck equation describing both
diffusion and dissipation. Let us look at the result in detail in
order to see why it works.  As already mentioned~\eqref{eq:3}
characterizes the bounded mappings giving rise to
translation-covariant quantum-dynamical semigroups. Allowing for
unbounded operators a further contribution becomes relevant
\begin{align}
   \label{eq:35}
   \mathcal{L}[{\hat \rho}]
=&-{i \over \hbar}
        \left[{\hat{\mathsf{y}}}_0+
        H_{\mathrm{\scriptscriptstyle eff}} ({\hat{\mathsf{x}}},{\hat{\mathsf{p}}})
        ,{\hat \rho}
        \right]
\\ \nonumber
&+\sum_{k=1}^{r}
\left[K_k{\hat \rho}K_k^{\dagger} -\frac{1}{2}\left\{K_k^{\dagger}K_k,{\hat \rho}\right\} \right],
\end{align}
where
\begin{align*}
K_k &={\hat{\mathsf{y}}}_k+L_k ({\hat{\mathsf{p}}}) ,
\\
 {\hat{\mathsf{y}}}_k &=\sum_{i=1}^{3}a_{ki}{\hat{\mathsf{x}}}_i \quad
  k=0,\ldots, r\leq 3 \quad a_{ki}\in \mathbb{R},
\\
H_{\mathrm{\scriptscriptstyle eff}}
({\hat{\mathsf{x}}},{\hat{\mathsf{p}}})&=\frac{\hbar}{2i}\sum_{k=1}^{r}
({\hat{\mathsf{y}}}_k L_k ({\hat{\mathsf{p}}}) -L_k^{\dagger} ({\hat{\mathsf{p}}}){\hat{\mathsf{y}}}_k),
\end{align*}
a particular realization of which leads to the model considered
in\cite{art11} (here considered in three dimensions with sum
over Cartesian indices understood)
\begin{equation}
   \label{eq:36}
   \mathcal{L}[{\hat \rho}]=- 
\frac{\bar{\lambda}}{2}\, \left[{\hat{\mathsf{x}}}, \left[ {\hat{\mathsf{x}}},
{\hat \rho} \right]
\right] - \frac{\bar{\lambda} \bar{\alpha}^2}{2\hbar^2}\, \left[{\hat{\mathsf{p}}}, \left[ {\hat{\mathsf{p}}}, {\hat \rho}
\right] \right]\, - \, i\, \frac{\bar{\lambda} \bar{\alpha}}{\hbar}\, \left[ {\hat{\mathsf{x}}},
\left\{ {\hat{\mathsf{p}}}, {\hat \rho} \right\} \right].
\end{equation}
The model is defined in terms of two constants $\bar{\lambda}$ and $\bar{\alpha}$,
assumed to vary with the mass of the particle as follows:
\begin{equation} \label{eqp}
\bar{\lambda} \; = \; \frac{m}{m_{0}}\, \bar{\lambda}_{0} \qquad\qquad \bar{\alpha}
\; = \; \frac{m_{0}}{m}\, \bar{\alpha}_{0},
\end{equation}
where $m_0$ is a reference mass, while $\bar{\lambda}_0$ and $\bar{\alpha}_0$ are fixed
constants. 
This scaling will turn out to be crucial in order to allow for the amplification mechanism.
Let us now directly check this mechanism,
considering a system of $N$ particles and taking the partial trace
with respect to the relative
coordinates. Using~\eqref{eq:21}, \eqref{eq:22}, \eqref{eq:23} and \eqref{eq:24} together with
\begin{equation}
   \label{eq:37}
   \bm{\pi}_i=\sum_{k=1}^{N}\Lambda_{ki}^{-1}\bm{p}_k
\end{equation}
with $\bm{\pi}_i$ the variables canonically conjugated to $\bm{r}_i$,
exploiting $\Lambda_{i1}^{-1}=1$ so that $\bm{\pi}_1=\bm{P}$ and therefore
\begin{equation}
   \label{eq:38}
   \bm{p}_i=\frac{m_i}{M}\bm{P}+\sum_{k=2}^{N}\Lambda_{ki}\bm{\pi}_k
\end{equation}
with $\bm{P}\equiv\sum_{i=1}^{N}\bm{p}_i$ the total momentum we obtain
as a consequence the important relation
\begin{equation}
   \label{eq:39}
   \sum_{i=1}^{N}\sum_{k=2}^{N}\Lambda_{ki}\bm{\pi}_k=0.
\end{equation}
Let us now take the partial trace of \eqref{eq:36} generalized to a
sum of $N$ contributions corresponding to particles of mass $m_i$ with
respect to the relative coordinates. One immediately has, exploiting
the linearity of commutator and anticommutator with respect to their
arguments, as well as the invariance of the trace operation under a
cyclic transformation
\begin{align}
   \label{eq:40}
   \mathcal{L}[\hat \rho_{\rm \scriptscriptstyle CM}]=
&
-\frac{1}{2}\sum_{i=1}^{N}
\bar{\lambda}_{i}
\Tr_{\rm \scriptscriptstyle rel}
\left(
\left [
\hat{\mathsf{X}}+\sum_{k=2}^{N}\Lambda_{ik}^{-1}\hat{\mathsf{r}}_k,
\left [
\hat{\mathsf{X}}+\sum_{k=2}^{N}\Lambda_{ik}^{-1}\hat{\mathsf{r}}_k,
\hat \rho_{\rm \scriptscriptstyle tot}
\right]
\right]
\right)      
\nonumber \\
&
\nonumber
-\frac{1}{2\hbar^2}\sum_{i=1}^{N}
\bar{\lambda}_{i}\bar{\alpha}_{i}^2
\Tr_{\rm \scriptscriptstyle rel}
\left(
\left [
\frac{m_i}{M}\hat{\mathsf{P}}+\sum_{k=2}^{N}\Lambda_{ki}\hat{\mathsf{\pi}}_k,
\left [
\frac{m_i}{M}\hat{\mathsf{P}}+\sum_{k=2}^{N}\Lambda_{ki}\hat{\mathsf{\pi}}_k,
\hat \rho_{\rm \scriptscriptstyle tot}
\right]
\right]
\right)      
\\
&
\nonumber
-\frac{i}{\hbar}\sum_{i=1}^{N}
\bar{\lambda}_{i}\bar{\alpha}_{i}
\Tr_{\rm \scriptscriptstyle rel}
\left(
\left [
\hat{\mathsf{X}}+\sum_{k=2}^{N}\Lambda_{ik}^{-1}\hat{\mathsf{r}}_k,
\left \{
\frac{m_i}{M}\hat{\mathsf{P}}+\sum_{k=2}^{N}\Lambda_{ki}\hat{\mathsf{\pi}}_k,
\hat \rho_{\rm \scriptscriptstyle tot}
\right\}
\right]
\right)      
\\
=&
\nonumber
-\frac{1}{2}\sum_{i=1}^{N}
\bar{\lambda}_{i}
\left [
\hat{\mathsf{X}},
\left [
\hat{\mathsf{X}},
\hat \rho_{\rm \scriptscriptstyle CM}
\right]
\right]
-\frac{1}{2\hbar^2}\sum_{i=1}^{N}
\bar{\lambda}_{i}\bar{\alpha}_{i}^2
\left(\frac{m_i}{M}\right)^2     
\left [
\hat{\mathsf{P}} ,
\left [
 \hat{\mathsf{P}} ,
\hat \rho_{\rm \scriptscriptstyle CM}
\right]
\right]
-\frac{i}{\hbar}\sum_{i=1}^{N}
\bar{\lambda}_{i}\bar{\alpha}_{i}\frac{m_i}{M}
\left [
\hat{\mathsf{X}},
\left \{
\hat{\mathsf{P}},
\hat \rho_{\rm \scriptscriptstyle CM}
\right\}
\right]
\\
&
-\frac{i}{\hbar}\sum_{i=1}^{N}
\bar{\lambda}_{i}\bar{\alpha}_{i}
\left [
\hat{\mathsf{X}},
\Tr_{\rm \scriptscriptstyle rel}
\left(
\left \{
\sum_{k=2}^{N}\Lambda_{ki}\hat{\mathsf{\pi}}_k,
\hat \rho_{\rm \scriptscriptstyle tot}
\right\}
\right)
\right].
\end{align}
Now the scalings given in~\eqref{eqp} show their relevance in leading to
\begin{align}
   \label{eq:42}
      \mathcal{L}[\hat \rho_{\rm \scriptscriptstyle CM}]=
&
-\frac{1}{2}
\bar{\lambda}_{\rm \scriptscriptstyle CM}
\left [
\hat{\mathsf{X}},
\left [
\hat{\mathsf{X}},
\hat \rho_{\rm \scriptscriptstyle CM}
\right]
\right]
-\frac{1}{2\hbar^2}
\bar{\lambda}_{\rm \scriptscriptstyle CM}\bar{\alpha}_{\rm \scriptscriptstyle CM}^2
\left [
\hat{\mathsf{P}} ,
\left [
 \hat{\mathsf{P}} ,
\hat \rho_{\rm \scriptscriptstyle CM}
\right]
\right]
-\frac{i}{\hbar}
\bar{\lambda}_{\rm \scriptscriptstyle CM}\bar{\alpha}_{\rm \scriptscriptstyle CM}
\left [
\hat{\mathsf{X}},
\left \{
\hat{\mathsf{P}},
\hat \rho_{\rm \scriptscriptstyle CM}
\right\}
\right]
\nonumber
\\
&
-\frac{i}{\hbar}
\bar{\lambda}_{\rm \scriptscriptstyle CM}\bar{\alpha}_{\rm \scriptscriptstyle CM}
\left [
{\hat{\mathsf{X}},
\Tr_{\rm \scriptscriptstyle rel}
\left(\left \{
\sum_{i=1}^{N}\sum_{k=2}^{N}\Lambda_{ki}\hat{\mathsf{\pi}}_k,
\hat \rho_{\rm \scriptscriptstyle tot}
\right\}
\right)  }    
\right]
\end{align}
where
\begin{equation}
   \label{eq:44}
   \bar{\lambda}_{\rm \scriptscriptstyle CM}=\frac{M}{m_{0}}\, \bar{\lambda}_{0} \qquad\qquad \bar{\alpha}_{\rm \scriptscriptstyle CM}=\frac{m_{0}}{M}\, \bar{\alpha}_{0},
\end{equation}
and thanks to~\eqref{eq:39} finally the result
\begin{equation}
   \label{eq:43}
         \mathcal{L}[\hat \rho_{\rm \scriptscriptstyle CM}]=
-\frac{1}{2}
\bar{\lambda}_{\rm \scriptscriptstyle CM}
\left [
\hat{\mathsf{X}},
\left [
\hat{\mathsf{X}},
\hat \rho_{\rm \scriptscriptstyle CM}
\right]
\right]
-\frac{1}{2\hbar^2}
\bar{\lambda}_{\rm \scriptscriptstyle CM}\bar{\alpha}_{\rm \scriptscriptstyle CM}^2
\left [
\hat{\mathsf{P}} ,
\left [
 \hat{\mathsf{P}} ,
\hat \rho_{\rm \scriptscriptstyle CM}
\right]
\right]
-\frac{i}{\hbar}
\bar{\lambda}_{\rm \scriptscriptstyle CM}\bar{\alpha}_{\rm \scriptscriptstyle CM}
\left [
\hat{\mathsf{X}},
\left \{
\hat{\mathsf{P}},
\hat \rho_{\rm \scriptscriptstyle CM}
\right\}
\right],
\end{equation}
reflecting the amplification mechanism. This is however just true
because the last term of~\eqref{eq:36} describing friction is simply
linear in the momentum and once the scalings~\eqref{eqp} are given one can exploit the fundamental
relation~\eqref{eq:39} to show that the term breaking the
amplification mechanism is zero. As it appears an exceptional
situation. A further possibility to allow for a momentum dependence in
the $L_j$ which allows for an explicit verification of the
amplification mechanism is something like $L_j\propto
e^{{-\frac{i}{\hbar}\bm{a}\cdot{\hat{\mathsf{p}}}}}$, as in the
structure of Weyl-covariant generators of quantum-dynamical
semigroups\cite{HolevoRMP33}, in such a case however one also has
boost covariance and therefore no friction effect leading to energy relaxation.

\section{Conclusions and outlook}
\label{sec:conclusions-outlook}

Exploiting the fact that the GRW master-equation for the description of
spontaneous localization has the property of being covariant under
translations, thus not breaking homogeneity of space,  it becomes
natural to write it in a way which makes the connection with the
general structure of translation-covariant master-equations obtained
by Holevo\cite{HolevoRMP32,HolevoRMP33} immediately apparent. In such
a way the formal connection with master-equations for the description
of decoherence\cite{art12} becomes straightforward and can be spelled
out in detail. This analysis in particular shows that the GRW
master-equation arises in a most natural way: it is essentially fixed
by asking for a Markovian dynamics to be described in terms of
momentum transfer events, and therefore translationally invariant,
where the momentum transfer in each event is random and described by a
Gaussian distribution.

One can further check that the amplification mechanism generally holds
for this class of translation-covariant master-equations for the
description of decoherence, and also specify a collisional decoherence
model which is formally exactly equivalent to the GRW master-equation.
Of course the parameters appearing in the model, related to bath
properties, have quite different orders of magnitudes, further
corroborating the known fact that decoherence is generally much
stronger than the GRW effect. Building on the formal analogy one can
ask the question whether the known extensions of decoherence model to
cope with dissipative effects\cite{art10,HornbergerPRL06} might be of
help in guessing a generalization of the GRW master-equation not
leading to the well--known infinite energy growth.  We show on the
basis of the general characterization of translation-covariant
master-equations available that the answer is generally negative, due
to the loss of a simple amplification mechanism. A notable exception
worked out in\cite{art11} is however pointed out, which presently
appears as the only possibility in this direction.  Note that the
problem of energy growth can also be overcome by considering different
models of quantum state reduction, where the localization operator is
given by the energy of the system, however such models do not
automatically grant space localization of macroscopic objects
(see\cite{AdlerJPA01} or\cite{Adler}, Chap. 6 for a general review of
the subject and the comparison between the two different approaches).

As a final remark we want to stress the different meaning of dynamical
reduction models and of the decoherence approach, the latter one not
providing a solution to the measurement
problem\cite{JoosLNPH99-AnastopoulosIJTP02- AdlerSHPMP03}. However the
formal analogies among the two research fields have been used in the
present paper, exploiting results on the structure of master-equations
for the description of decoherence in order to better understand
properties of dynamical reduction models.

\section*{Acknowledgements}
This work was partially supported by MIUR under
PRIN05 and FIRB. The author is grateful to Prof. L. Lanz and Dr. A. Bassi for useful
discussions on the subject of the paper and careful reading of the manuscript.

\end{document}